\documentclass{article}

\usepackage{microtype}
\usepackage{graphicx}
\usepackage{subfigure}
\usepackage{booktabs} %

\usepackage{hyperref}

\usepackage[preprint]{icml2019}

\usepackage[T1]{fontenc}
\usepackage{tgtermes}
\usepackage{amssymb}
\newcommand{\mathbold}[1]{\ensuremath{\boldsymbol{\mathbf{#1}}}}

\usepackage{amsmath}

\usepackage[english]{babel}
\usepackage[parfill]{parskip}
\usepackage{afterpage}
\usepackage{enumitem}
\usepackage{framed}
\usepackage{xspace}

\usepackage{lineno}

\newcounter{parcount}

\usepackage[usenames,dvipsnames]{xcolor}
\definecolor{shadecolor}{gray}{0.9}

\usepackage{booktabs, array}

\usepackage[nameinlink]{cleveref}
\creflabelformat{equation}{#2\textup{#1}#3}  %

\usepackage
[acronym,smallcaps,nowarn,section,nogroupskip,nonumberlist]{glossaries}
\glsdisablehyper{}

\newacronym{VI}{vi}{variational inference}
\newacronym{KL}{kl}{Kullback-Leibler}
\newacronym{ELBO}{elbo}{\emph{evidence lower bound}}
\newacronym{MCMC}{mcmc}{Markov chain Monte Carlo}

\newcommand{\Gam}{\textrm{Gam}}

\newcommand{\N}{\mathcal{N}}
\newcommand{\Bern}{\textrm{Bern}}

\newcommand{\partiald}[2]{\frac{\partial #1}{\partial #2}}

\newcommand{\mbbeta}{\mathbold{\beta}}

\newcommand{\mblambda}{\mathbold{\lambda}}

\newcommand{\mbzero}{\mathbold{0}}

\usepackage{tikz}

\pgfdeclarelayer{edgelayer}
\pgfdeclarelayer{nodelayer}
\pgfsetlayers{edgelayer,nodelayer,main}

\definecolor{hexcolor0xbfbfbf}{rgb}{0.749,0.749,0.749}

\tikzset{>=latex}
\tikzstyle{none}   = [inner sep=0pt]
\tikzstyle{line}   = [ -, thick, shorten <=1pt, shorten >=1pt ]
\tikzstyle{arrow}  = [ ->, thick, shorten <=1pt, shorten >=1pt ]
\tikzstyle{ardash} = [ dashed, ->, thick, shorten <=1pt, shorten >=1pt ]

\tikzstyle{empty}=[circle,opacity=0.0,text opacity=1.0,inner sep=0pt]
\tikzstyle{box}=[rectangle,fill=White,draw=Black]
\tikzstyle{filled}=[circle,thick,fill=hexcolor0xbfbfbf,draw=Black]
\tikzstyle{hollow}=[circle,thick,fill=White,draw=Black]
\tikzstyle{param}=[rectangle,fill=Black,draw=Black,inner sep=0pt,minimum width=4pt,minimum height=4pt]
\tikzstyle{paramhollow}=[rectangle,thick,fill=White,draw=Black,inner sep=0pt,minimum
width=4pt,minimum height=4pt]

\usepackage{pgfplots}                               %
\pgfplotsset{compat=newest}
\pgfplotsset{plot coordinates/math parser=false}
\newlength\figureheight
\newlength\figurewidth
\newlength\figureheightsmall
\newlength\figurewidthsmall
\definecolor{POSTcolor}{rgb}{0.48, 0.20, 0.58} %
\definecolor{Qcolor}{rgb}{0.00, 0.53, 0.22} %

\def \I { {\mbox{I}}}

\icmltitlerunning{NeuTra-lizing Bad Geometry in Hamiltonian Monte Carlo Using Neural Transport}

\begin{document}

\twocolumn[
\icmltitle{NeuTra-lizing Bad Geometry in Hamiltonian Monte Carlo Using Neural Transport}

\icmlsetsymbol{equal}{*}

\begin{icmlauthorlist}
\icmlauthor{Matthew Hoffman}{equal,goo}
\icmlauthor{Pavel Sountsov}{equal,goo}
\icmlauthor{Joshua V. Dillon}{goo}
\icmlauthor{Ian Langmore}{goo}
\icmlauthor{Dustin Tran}{goo}
\icmlauthor{Srinivas Vasudevan}{goo}
\end{icmlauthorlist}

\icmlaffiliation{goo}{Google Research}

\icmlcorrespondingauthor{Matthew Hoffman}{mhoffman@google.com}

\icmlkeywords{Monte Carlo Methods, Deep Generative Models, Bayesian Methods}

\vskip 0.3in
]

\printAffiliationsAndNotice{\icmlEqualContribution} %

\begin{abstract}
    Hamiltonian Monte Carlo is a powerful algorithm for sampling from
    difficult-to-normalize posterior distributions. However, when the geometry
    of the posterior is unfavorable, it may take many expensive evaluations of
    the target distribution and its gradient to converge and mix. We propose
    neural transport (NeuTra) HMC, a technique for learning to correct this
    sort of unfavorable geometry using inverse autoregressive flows (IAF), a
    powerful neural variational inference technique. The IAF is trained to
    minimize the KL divergence from an isotropic Gaussian to the warped
    posterior, and then HMC sampling is performed in the warped space. We
    evaluate NeuTra HMC on a variety of synthetic and real problems, and find
    that it significantly outperforms vanilla HMC both in time to reach the
    stationary distribution and asymptotic effective-sample-size rates.
\end{abstract}

\section{Introduction}

Markov chain Monte Carlo (MCMC) is a powerful meta-strategy
for sampling from unnormalized distributions. One sets up a Markov
chain with the desired stationary distribution, and simulates it to
generate correlated samples from that distribution. MCMC's great strength
is that, given enough computation (and subject to mild
ergodicity conditions), it is guaranteed to generate samples from the
target distribution. However, if successive
samples from the chain are highly correlated, then the chain will take
a long time to produce independent samples.

Hamiltonian Monte Carlo \citep[HMC; ][]{duane1987hybrid,neal2011mcmc}
is an MCMC algorithm that is particularly well suited to sampling from
high-dimensional continuous distributions. It introduces a set of auxiliary
variables that let one generate Metropolis-Hastings proposals
\citep{metropolis1953equation,hastings1970mh} by simulating the dynamics of a
fictional Hamiltonian physical system. However, HMC is not a silver
bullet. When the geometry of the target distribution is unfavorable,
it may take many evaluations of the log-probability of the target
distribution and its gradient for the chain to mix between faraway
states \citep{betancourt2017conceptual}.

\citet{parno2014transport} proposed a way to fix such unfavorable
geometry by applying a reversible transformation (or ``transport map'') that
warps the space in which the chain is simulated.
If we are interested
in sampling from a distribution $p(\theta)$ over a real-valued vector
$\theta$, then we can equivalently apply a bijective change of
variables $z=f^{-1}(\theta)$ and sample from
$p(z)=p(\theta=f(z))|\partiald{f}{z}|$. If $f$ is chosen so that
the geometry of $p(z)$ is amenable to efficient MCMC sampling (for example,
if $p(z)=\N(z; \mbzero, \I)$), then one can run an MCMC chain in $z$-space
and then push the $z$ samples forward through $f$ to get samples from
$p(\theta)$.

The question then becomes: what family of transformations $f$ should
we use, and how do we find the best member of that family?
\citet{titsias2017learning} proposes using a diagonal affine transformation,
which can be insufficiently powerful. \citet{parno2014transport} and
\citet{marzouk2016introduction} proposed an $f$ based on a series of polynomial
regressions that minimize the forward or reverse Kullback-Leibler (KL)
divergence between $p(z)$ and $\N(z; \mbzero, \I)$. Unfortunately this approach
is too expensive to use in high dimensional problems.

In this work, we propose using a transport map consisting of a series of inverse
autoregressive flows \citep[IAFs; ][]{kingma2016iaf} parameterized by
neural networks fit using variational inference.

Our main contributions are:
\begin{itemize}
\item We improve on the transport-map MCMC approach of
  \citet{marzouk2016introduction} by using more powerful and scalable
  IAF maps, and by using the more powerful gradient-based HMC sampler.
\item We adapt this strategy to train variational autoencoders
  \citep{rezende2014stochastic,kingma2014autoencoding}.
\item We evaluate our neural-transport HMC (NeuTra HMC for short)
  approach on a variety of synthetic and real problems, and find that
  it can consistently outperform HMC, often by an order of magnitude.
\end{itemize}

\section{Neural Transport MCMC}
\label{sec:neutra}

In order to describe NeuTra MCMC, we first outline its two main ingredients:
Hamiltonian Monte Carlo, a gradient-based MCMC algorithm for sampling from a
target distribution; and normalizing flows, which are reversible
transformations that warp simple distributions so that they approximate complex ones.

\subsection{Hamiltonian Monte Carlo}
\label{sub:hamiltonian}

Hamiltonian Monte Carlo \citep[HMC; ][]{duane1987hybrid,neal2011mcmc} is a
Markov chain Monte Carlo algorithm that introduces an auxiliary momentum
variable $m_d$ for each parameter $\theta_d$ in the state space to transition
over. These momentum variables follow a multivariate normal distribution
(typically with identity covariance). The augmented, unnormalized joint
distribution is
\begin{equation*}
\textstyle p(\theta, m) \propto \exp\{\mathcal{L}(\theta) - \frac{1}{2}m^\top m\},
\end{equation*}
where $\mathcal{L}(\theta)$ is the log-probability of the variables of interest
$\theta$ (up to a normalizing constant).
Intuitively, the augmented model acts as a fictitious Hamiltonian system where
$\theta$ represents a particle's position, $m$ represents the particle's momentum,
$\mathcal{L}(\theta)$ is the particle's negative
potential energy, $m^\top m/2$ is the particle's kinetic energy, and $\log
p(\theta, m)$ is the total negative energy of the particle.

We simulate the system's Hamiltonian dynamics using the leapfrog integrator,
which applies the updates
\begin{equation}
\begin{split}
\label{eq:leapfrog}
m^{t + \epsilon / 2} &= m^t + (\epsilon/2) \nabla_\theta \mathcal{L}(\theta^t),
\\
\theta^{t + \epsilon} &= \theta^t + \epsilon m^{t + \epsilon / 2},
\\
m^{t + \epsilon} &= m^{t + \epsilon / 2} + (\epsilon/2) \nabla_\theta \mathcal{L}(\theta^{t + \epsilon}),
\end{split}
\end{equation}
where superscripts are time indices. The updates for each coordinate are additive
and depend only on the other coordinates, which implies the leapfrog integrator is
reversible and conserves volume.

Each HMC update proceeds by first resampling momentum variables
$m\sim \N(\mbzero, \I)$. It
then applies $L$ leapfrog updates to the position and momentum
$(\theta, m)$, generating a new state $(\tilde{\theta},
\tilde{m})$. The state $(\tilde{\theta}, -\tilde{m})$ is proposed, and
it is accepted or rejected according to the Metropolis algorithm
with probability $\min\{1, \frac{p(\tilde\theta, \tilde m)}{p(\theta, m)}\}$
\citep{metropolis1953equation}.
Since Hamiltonian dynamics conserve total energy, if the leapfrog
discretization is accurate the total change in energy
$\log p(\theta, m) - \log p(\tilde \theta, \tilde m)$ will be small,
and the proposal will probably be accepted.

The leapfrog integrator is accurate to $O(\epsilon^2)$; the acceptance
rate can be kept high by making $\epsilon$ sufficiently small. But if
$\epsilon$ is reduced, then the number of leapfrog steps $L$ must
increase accordingly to keep the total distance traveled roughly
constant, which is expensive insofar as HMC's cost per iteration is
typically dominated by the gradient computation in
\Cref{eq:leapfrog}. For some target distributions with unfavorable
geometry, there will be a mix of ``stiff'' directions with high
curvature (requiring a small $\epsilon$) and less-constrained
directions with low curvature (requiring many leapfrog steps to
explore). Even worse, if the target distribution has tails that are
either too heavy or too light, HMC may mix arbitrarily slowly
regardless of how many leapfrog steps are applied per iteration
\citep{livingstone2016geometric}.
On the other hand, if the bulk of the target distribution is strongly
log-concave then HMC can mix very efficiently \citep{mangoubi2017rapid}.
In summary, we should expect HMC to be most efficient when applied to
roughly isotropic distributions with roughly Gaussian tail behavior.

\subsection{Normalizing Flows and Variational Inference}
\label{sub:reversible}

Let $\theta=f_\phi(z)$ for a bijective, continuously differentiable
function $f_\phi$ parameterized by some vector $\phi$. If $z$ has some
distribution $q(z)$, then the standard change-of-variables identity
states that
\begin{equation}
\begin{split}
\textstyle q(\theta)=q(z)|\partiald{f}{z}|^{-1}.
\end{split}
\end{equation}
If we want to make $q(\theta)$ approximate some target distribution
$p(\theta)\propto e^{-A}\pi(\theta)$, we can tune $\phi$ to maximize
the evidence lower bound (ELBO) \citep{rezende2015normalizing}:
\begin{equation}
\begin{split}
\label{eq:elbo}
\textstyle
\mathcal{L}(\phi) &=
A - \mathrm{KL}(q(\theta)\mid\mid p(\theta))
\\ &=
\textstyle
\int_\theta q(\theta) \log\frac{\pi(\theta)}{q(\theta)} d\theta
\\ &=
\textstyle
\int_z q(z) \log\frac{\pi(f(z))}{q(z)|\partiald{f}{z}|^{-1}} dz.
\end{split}
\end{equation}
If we can sample from $q(z)$ and compute 1) $q(z)$, 2) the
log-determinant of the Jacobian $|\partiald{f}{z}|^{-1}$, and 3) the
unnormalized density $\pi(\theta)$, then we can compute an unbiased
Monte Carlo estimate of the ELBO (and, using automatic differentiation,
its derivative) by evaluating the
log-ratio in \Cref{eq:elbo} at a $z$ sampled from $q(z)$. We can use
these estimates to maximize the ELBO w.r.t. $\phi$, and therefore
minimize the KL divergence from $q(\theta)$ to $p(\theta)$.

Even if $q(z)$ is a simple distribution ($q(z)=\N(z; \mbzero, \I)$ is a common
choice), a sufficiently powerful flow $f_\phi$ can
transform it into a close approximation to $p(\theta)$. More
expressive maps can be achieved by stacking multiple simpler maps, since each
map is invertible and the overall Jacobian is the product of the individual map
Jacobians:
$\partiald{f(f'(z))}{z}=\partiald{f}{f'}\partiald{f'}{z}$.

\emph{Inverse autoregressive flows} \citep[IAF; ][]{kingma2016iaf} are
a powerful, efficient class of normalizing flows parameterized by
neural networks. The idea is to construct $f$ such that
\begin{equation}
\begin{split}
\label{eq:iaf}
f_i(z) = z_i \sigma_i(z_{1:i-1}; \phi) + \mu_i(z_{1:i-1}; \phi),
\end{split}
\end{equation}
that is, $\theta_i$ is a shifted and scaled version of $z_i$, where
the shift and scale are parameterized by a neural network. The
transformation allows each output dimension to depend on previous input
dimensions using arbitrary (possibly noninvertible) neural networks, and the
mapping can be computed in parallel across $i$. In addition, the Jacobian
$\partiald{f}{z}$ is lower triangular by construction, so its determinant is
simply $\prod_i \sigma_i$.

\subsection{Neural-Transport MCMC}
\label{sub:neutra}

\citet{marzouk2016introduction} note that the process of fitting a
transport map by variational inference can be interpreted in terms
of the inverse map. KL divergence is invariant to changes of
variables, so minimizing $\mathrm{KL}(q(\theta)\mid\mid p(\theta))$,
is equivalent to minimizing $\mathrm{KL}(q(z)\mid\mid p(z))$. That is,
in $z$-space variational inference is trying to warp the
pulled-back target distribution $p(z)$ to look as much as possible like
the fixed distribution $q(z)$.

If we have tuned the parameters $\phi$ of the map $f_\phi$ so that
$p(z)=p(\theta=f_\phi(z))|\partiald{f}{z}|\approx q(z)$, and $q(z)$ is
relatively easy to sample from by MCMC (for example, because it is a
simple distribution such as an isotropic Gaussian), then we can
efficiently sample from $p(\theta)$ by running a Markov chain whose
target distribution is $p(z)$.

We can think of this procedure in either of two ways: on the one hand,
we are using MCMC to correct for the failure of variational inference
to make $q(\theta)$ exactly match $p(\theta)$. On the other,
we are using the information that $q(\theta)$ has learned about
$p(\theta)$ to accelerate mixing of our MCMC algorithm of choice.

\citet{marzouk2016introduction} proposed using maps based on a series
of polynomial approximations. These worked reasonably well in the
low-dimensional inverse problems they considered, but to apply them to
problems in even moderately high dimensions they had to resort to
stronger independence assumptions that lead to less flexible maps.

We propose two main improvements to the approaches of
\citet{marzouk2016introduction} that scale their transport-map MCMC
idea to the higher-dimensional problems common in Bayesian statistics
and probabilistic machine learning. First, we use Hamiltonian Monte
Carlo \citep[HMC; ][]{duane1987hybrid,neal2011mcmc} which is able to mix
dramatically faster than competing MCMC methods in high dimensions due to its
use of gradient information; on $D$-dimensional
strongly log-concave distributions, HMC can generate samples in
$\Theta(D^{1/4})$ \citep{mangoubi2017rapid}, dramatically faster than
gradient-free methods like random-walk Metropolis (which requires
$\Omega(D)$ steps). The results of \citet{mangoubi2017rapid} will
apply if we can find a map such that the bulk of the mass of the
transformed distribution $p(z)$ is in a region where $p(z)$ is
strongly log-concave (e.g., if $p(z)\approx\mathcal{N}(z; 0, I)$).
Second, we use IAFs, which are more scalable (and likely more
powerful) than polynomial maps. We call the resulting approach
neural-transport HMC, or NeuTra HMC for short.

To summarize, given a target distribution $p(\theta)$, NeuTra HMC
proceeds in three steps:

1. Fit an IAF to minimize the KL divergence between
$q(\theta)=q(z)|\partiald{f}{z}|^{-1}$ and $p(\theta)$.

2. Run HMC with target distribution
\\$p(z)=p(\theta=f(z))|\partiald{f}{z}|$, initialized with a sample
from $q(z)$.

3. Push the $z$-space samples forward through $f$ to get samples from
$p(\theta)$.

Note that we never need to compute the inverse transformation
$f^{-1}(\theta)$, which is expensive for IAFs.

\begin{figure}[!tb]
  \centering
    \includegraphics[width=1.1\columnwidth]{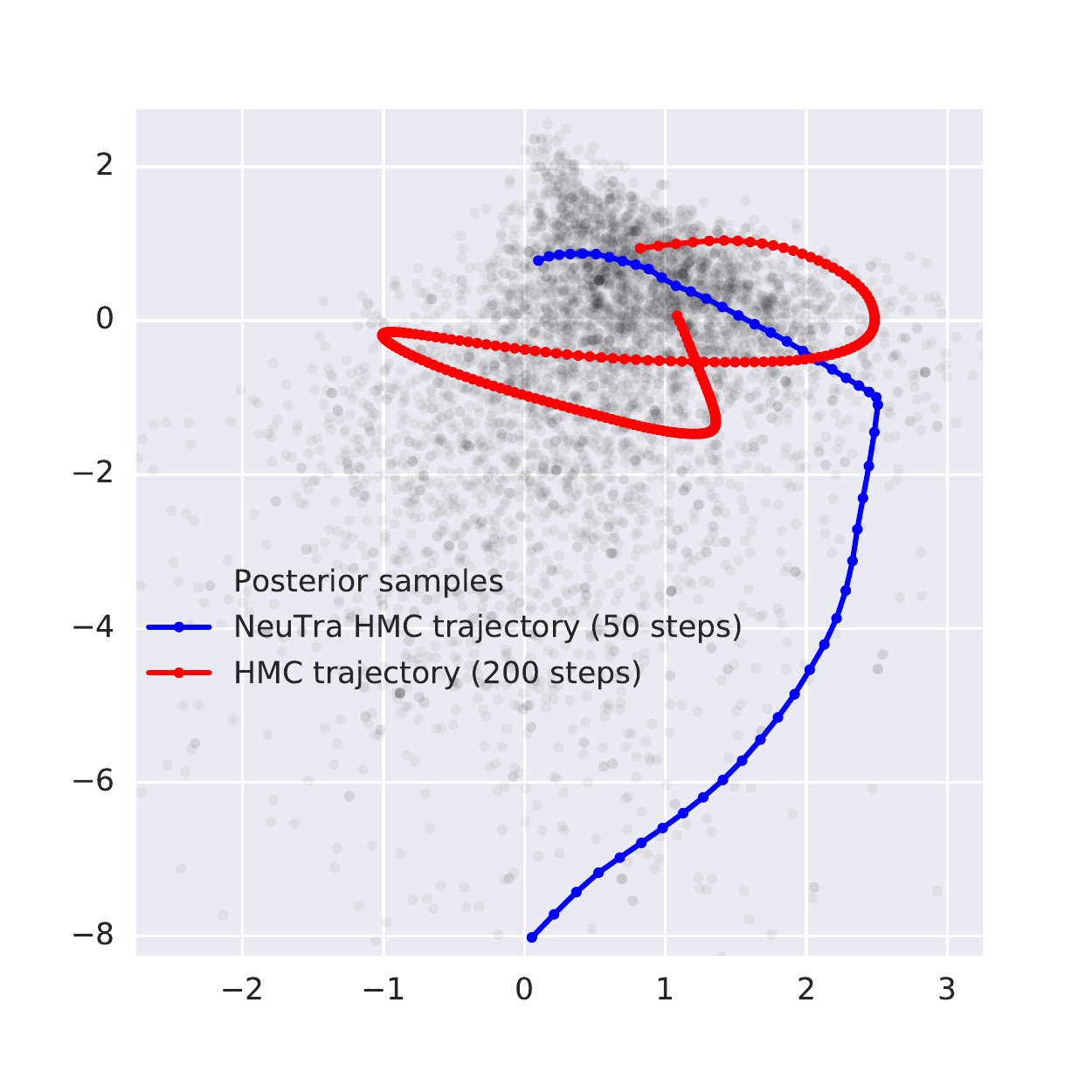}
\caption{
  2-dimensional projection of the trajectories
  obtained by running the leapfrog integrator in
  the original (red) and warped (blue) parameter spaces.
  Warped trajectories are displayed in the original parameter space.
  The negative potential energy is the log-posterior of
  the soft-sparsity logistic regression model from \Cref{sub:unconditional}.
  The warped space is defined by stacking three IAFs. Simulating the dynamics
  in the warped space, where the transformed potential energy function
  is simpler, leads to trajectories that move further in fewer steps.
  Note also that the NeuTra HMC's effective step size increases in the
  tails, where there is lower curvature.
}
\label{fig:trajectories}
\end{figure}

\Cref{fig:trajectories} illustrates how simulating Hamiltonian
dynamics in the $z$-space defined by an IAF can produce trajectories
that quickly explore $\theta$-space in relatively few steps.

\subsubsection{NeuTra in the amortized setting}
\label{subsub:amortized}
Amortized variational inference
\citep{kingma2014autoencoding,rezende2014stochastic,gershman2014amortized}
is a popular strategy for learning and inference in
latent-variable models. Rather than optimize the parameters
of a single variational distribution $q(\theta; \phi)$ to minimize the KL
divergence to a single posterior $p(\theta\mid x)$, one trains a
\emph{conditional} variational distribution $q(\theta\mid x; \phi)$,
typically parameterized by a neural network. The cost of fitting
$\phi$ can be \emph{amortized} over many values of
$x$.

IAFs and other neural-network-based transport maps are well suited to this sort
of strategy; one need only design the network to take some auxiliary inputs $x$
as well as the latent vector $z$, or have the base distribution be
parameterized by $x$. Since NeuTra HMC is agnostic to how the map was created,
it also works in the amortized setting.

\section{Related Work}
\label{sec:related}

NeuTra HMC has ties to several threads of related work. In this section,
we describe some of these connections.

\subsection{Riemannian Manifold HMC}
\label{subsection:rmhmc}
Riemannian manifold HMC \citep[RMHMC; ][]{girolami2011riemann} tries to
speed mixing by accounting for the information geometry of the
target distribution. Where HMC generates proposals by simulating
Hamiltonian dynamics in Euclidean space, RMHMC simulates Hamiltonian
dynamics in a Riemannian space with a position-dependent metric. When
this metric is chosen appropriately, RMHMC can make rapid
progress in very few steps.

Despite this, RMHMC has some significant downsides compared to
standard Euclidean HMC algorithms: since the RMHMC Hamiltonian is
non-separable, it requires a more complicated, expensive, and
sensitive implicit numerical integrator; the commonly used Fisher
metric must be derived by hand for each new model, and is not always
available in closed form; if the metric changes rapidly as a function
of position, then the integrator may still need to use a small step
size; and in high dimensions it may be expensive to compute the
metric, its derivatives, its inverse, and its determinant.

Below, we show that the continuous-time dynamics of NeuTra HMC are in fact
equivalent to those of RMHMC with a metric defined by the Jacobian
of the map. This suggests that NeuTra HMC may be able to achieve
many of the benefits of RMHMC with much lower implementation and
computational complexity. For example, \Cref{fig:trajectories} demonstrates
RMHMC-style locally adaptive step size behavior.

Let $J\triangleq \partiald{f}{z}$ and $\ell(\theta)\triangleq\log p(\theta)$.
The Hamiltonian defined by NeuTra HMC is
\begin{equation}
\begin{split}
\textstyle
H_\mathrm{NT} = -\ell(f(z)) - \log|J| + \frac{1}{2}m^\top m.
\end{split}
\end{equation}
Now, consider the non-separable Hamiltonian that arises if we work in
the original $\theta$-space and define a position-dependent metric
$G\triangleq(JJ^\top)^{-1}$:
\begin{equation}
  \label{equation:h-riemann}
\begin{split}
H_\mathrm{RM} = \textstyle-\ell(\theta) + \frac{1}{2}m'^\top G^{-1}m'
+ \frac{1}{2}\log |G|.
\end{split}
\end{equation}
Now, if we define $m' \triangleq (J^\top)^{-1} m$, then we see that
\begin{equation}
\begin{split}
H_\mathrm{NT} &=\textstyle -\ell(f(z)) - \log|J| + \frac{1}{2}m^\top m
\\ &\textstyle= -\ell(f(z)) - \frac{1}{2}\log|JJ^\top| +
\frac{1}{2}m'^\top J J^\top m'
\\ &= H_\mathrm{RM}.
\end{split}
\end{equation}
That is, the Hamiltonians are equivalent. Now, consider the dynamics
over $(\theta, m')$ implied by $H_\mathrm{NT}$:
\begin{equation}
\begin{split}
\textstyle\partiald{\theta}{t} &\textstyle= \partiald{f}{z}\partiald{z}{t}
= J\left(\partiald{H_\mathrm{NT}}{m}\right)^\top;
\\\textstyle \partiald{m'}{t} &\textstyle= \partiald{m'}{m}\partiald{m}{t}
= -(J^\top)^{-1}\left(\partiald{H_\mathrm{NT}}{z}\right)^\top.
\end{split}
\end{equation}
These are the same as the dynamics implied by $H_\mathrm{RM}$:
\begin{equation}
\begin{split}
\textstyle\partiald{\theta}{t} &\textstyle= \left(\partiald{H_\mathrm{RM}}{m'}\right)^\top
= \left(\partiald{H_\mathrm{RM}}{H_\mathrm{NT}}
\partiald{H_\mathrm{NT}}{m}\partiald{m}{m'}\right)^\top
= J\left(\partiald{H_\mathrm{NT}}{m}\right)^\top;
\\\textstyle \partiald{m'}{t} &\textstyle= -\left(\partiald{H_\mathrm{RM}}{\theta}\right)^\top
= -\left(\partiald{H_\mathrm{RM}}{H_\mathrm{NT}}
\partiald{H_\mathrm{NT}}{z}\partiald{f}{z}^{-1}\right)^\top
    \\ &\textstyle= -(J^\top)^{-1}\left(\partiald{H_\mathrm{NT}}{z}\right)^\top,
\end{split}
\end{equation}
where we use the fact that
$\partiald{H_\mathrm{RM}}{H_\mathrm{NT}} = 1$.

NeuTra HMC therefore has the potential to deliver speedups comparable
to RMHMC without the complications and expense mentioned above. The
main downside is that this potential will only be realized if we learn
a good explicit map $f$, whereas RMHMC's metric can be computed from
purely local information.

\subsection{Learned MCMC Kernels}

Classical adaptive MCMC algorithms \citep{andrieu2008tutorial} try to
tune parameters such as step sizes to achieve target acceptance rates
or maximize convergence speed \citep[e.g.; ][]{pasarica2010adaptively}. More
recently, \citet{levy2018generalizing} proposed L2HMC, an adaptive
MCMC algorithm to tune a generalization of the leapfrog integrator
parameterized by a neural network; like NeuTra, it aims to use
powerful models to speed up mixing, but the L2HMC integrator is not
symplectic, and therefore may sacrifice some of the leapfrog
integrator's stability over long trajectories
\citep{neal2011mcmc,betancourt2017conceptual}. \citet{song2017generative}
propose a different neural MCMC approach based on adversarial
training, although it lacks strong guarantees of convergence. Whereas
NeuTra uses a variational approximation to speed up MCMC,
\citet{li2017approximate} propose schemes for using MCMC to improve a
variational approximation.

\citet{neal2011mcmc} suggests choosing a covariance matrix for the
momenta in HMC based on an estimate of the covariance of the target
distribution (or its diagonal), and observes that this corresponds to
doing HMC with an identity covariance under a linear change of
variables. The Stan software package \citep{carpenter2017stan} adapts
this covariance matrix while sampling, effectively tuning a more efficient
parameterization. NeuTra goes beyond the linear case.

\subsection{Deep Generative Models}

A few papers in the deep generative modeling literature have proposed
hybrids of variational inference and MCMC.  Several
\citep[e.g.;
][]{salimans2015markov,zhang2018measure,caterini2018hamiltonian} have
considered using MCMC-based variational bounds to do approximate
maximum-likelihood training of variational autoencoders (VAEs).
\citet{hoffman2017learning} proposed using the standard deviations of
a mean-field Gaussian variational distribution to tune per-variable
HMC step sizes, which is equivalent to doing HMC under the linear
change of variables that makes the variational distribution a standard
normal \citep{neal2011mcmc}. \citet{titsias2017learning} proposes
a heuristic for training an MCMC transport map by maximizing the
log-density of the last sample in the chain; since this method ignores
the intractable entropy term bounded by \citet{salimans2015markov}, it
is not clear that it actually encourages mixing as opposed to
mode-finding.

\section{Experiments}

\begin{figure*}[!tb]
  \centering
  \includegraphics[width=0.26\textwidth]{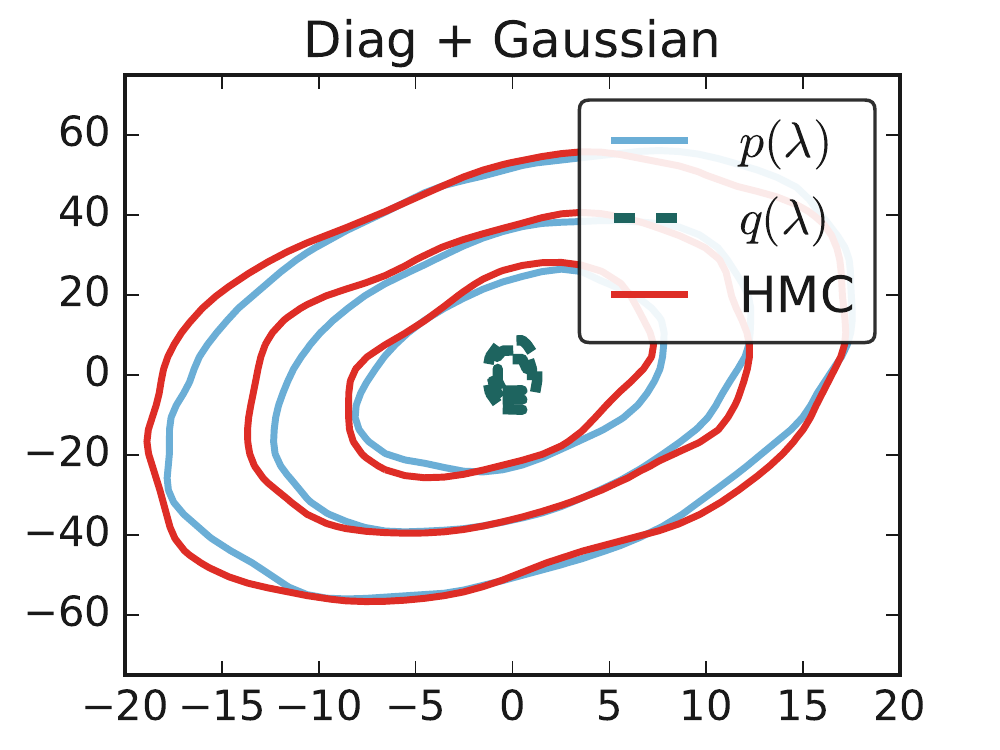}\hspace{-0.4cm}
  \includegraphics[width=0.26\textwidth]{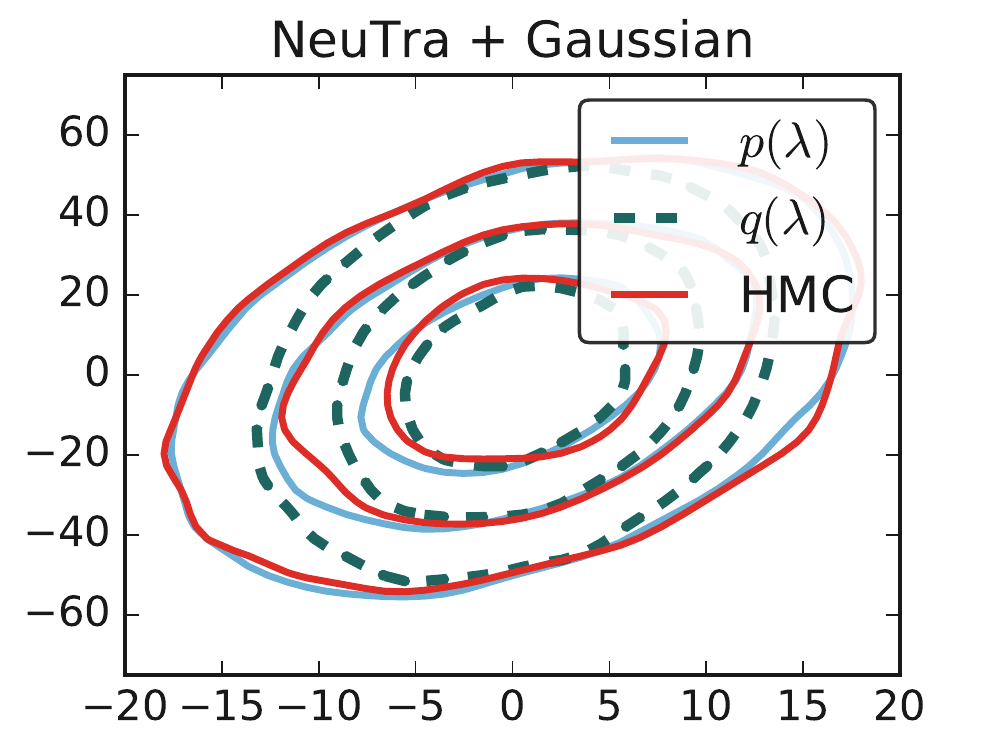}\hspace{-0.4cm}
  \includegraphics[width=0.26\textwidth]{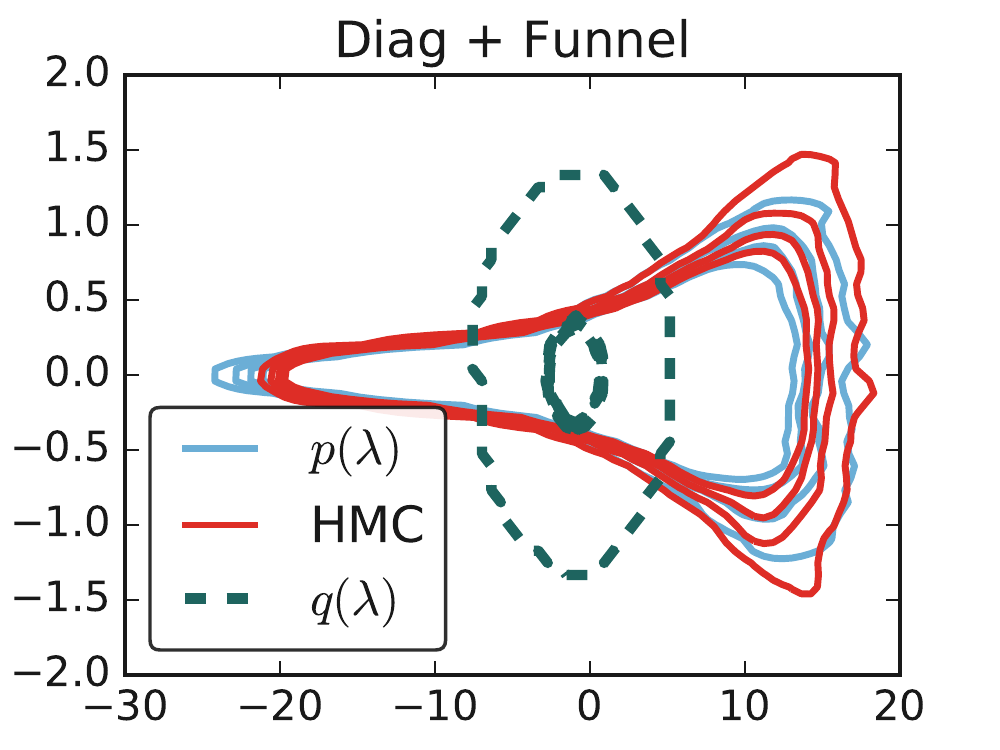}\hspace{-0.4cm}
  \includegraphics[width=0.26\textwidth]{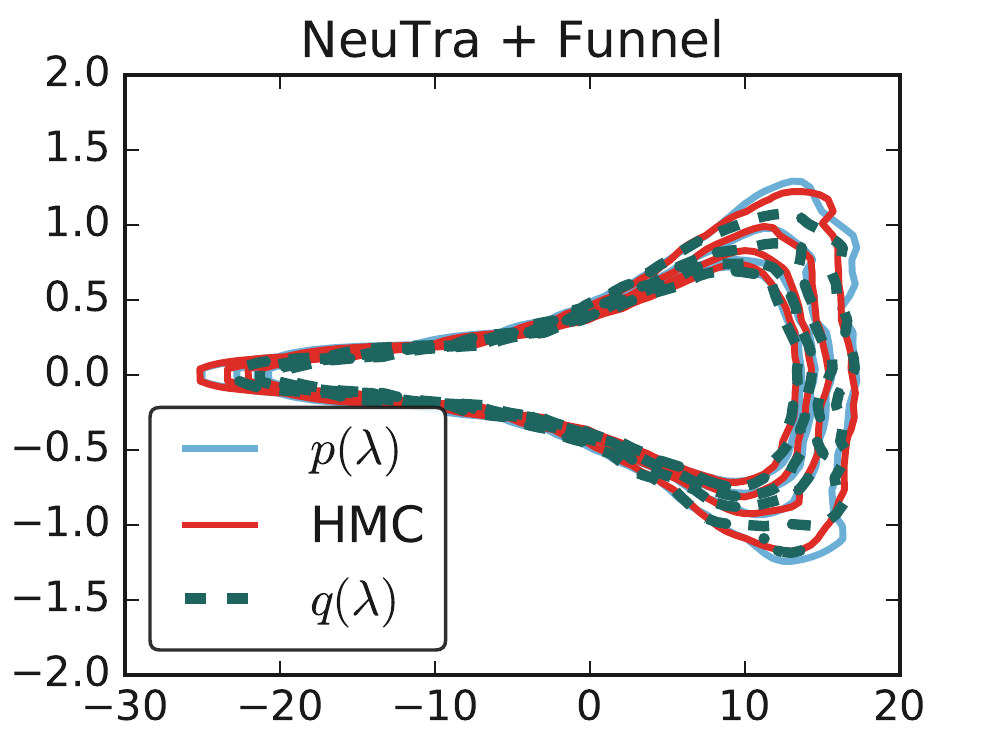}
\caption{
  2-dimensional projections of samples from an NeuTra IAF variational distribution,
  diagonal covariance gaussian variational distribution and the corresponding HMC samples on the two synthetic distributions.
  While the IAF is not always able to perfectly match the target distribution,
  HMC and NeuTra HMC are able to generate good samples (although HMC with the diagonal transport map has
  a little trouble with the neck of the funnel).
}
\label{fig:samples}
\end{figure*}

We evaluate NeuTra HMC's performance on four target distributions: two
synthetic problems, sparse logistic regression models
applied to the German credit dataset, and a variational autoencoder
applied to the MNIST dataset.  All experimental
code is open-sourced at
\url{https://github.com/google-research/google-research/tree/master/neutra}.

\subsection{Unconditional Target Distributions}
\label{sub:unconditional}

\paragraph{Ill-conditioned Gaussian:}
In order to test how samplers can handle a highly non-isotropic distribution,
we take a $D=100$ dimensional Gaussian distribution with the covariance matrix
with eigenvalues sampled from $\Gam(\alpha=0.5, \beta=1)$. The covariance matrix is
quenched (sampled once and shared among all the experiments). In practice, the
eigenvalues range over 6 orders of magnitude.

\paragraph{Neal's Funnel Distribution:}
We consider a $D=100$ dimensional version of the funnel distribution described by
\citet{neal2003slice}. This distribution mimics the geometry of a
hierarchical Bayesian prior with a centered parameterization, which is
known to be problematic for HMC \citep{neal2011mcmc}:
\begin{align*}
    p(\theta) = \N(\theta_0; 0, 1) \N(\theta_{1..99}; \mbzero, \I \exp ( 2\theta_0) ).
\end{align*}

\paragraph{Sparse logistic regression:}
As a non-synthetic example, we consider a hierarchical logistic
regression model with a sparse prior applied to the German credit dataset. We
use the numeric variant of the dataset, with the covariates standardized to
range between -1 and 1. With the addition of a constant factor, this yields 25
covariates.

The model is defined as follows:
\begin{equation}
\begin{split}
\tau &\sim \Gam(\alpha=0.5, \beta=0.5) \\
\lambda_d &\sim \Gam(\alpha=0.5, \beta=0.5) \\
\beta_d &\sim \N(0, 1) \\
y_n &\sim \Bern(\sigma(x_n^\top (\tau \mbbeta\circ \mblambda)))
\end{split}
\end{equation}
where $\Gam$ is the Gamma distribution, $\tau$ is the overall scale,
$\mblambda$ are per-dimension scales, $\mbbeta$ are the non-centered covariate weights,
$\mbbeta\circ \mblambda$ denotes the elementwise product of $\mbbeta$ and $\mblambda$, and
$\sigma$ is the sigmoid function. The sparse gamma prior on $\mblambda$
imposes a soft sparsity prior on the weights, which could be used for variable
selection. This parameterization uses D=51 dimensions. We log-transform
$\tau$ and $\mbbeta$ to make them unconstrained.

\subsubsection{Transport maps and training procedure}

For each distribution we consider IAFs with 2 hidden layers per
flow, three stacked flows, ELU nonlinearity \citep{clevert2015fast} and hidden dimensionality equal
to the target distribution dimensionality. 
When stacking multiple flows, we
reverse the order of dimensions between each flow. We also considered two
non-neural maps as baselines: a per-component scale vector (``Diag'') and shift
and a lower-triangular affine transformation (``Tril'') and shift. We use the
diagonal map as the baseline HMC method, as that approximates the standard
practice of basic preconditioning of that method. For sparse logistic
regression, we additionally scaled the base Gaussian distribution by 0.1 when
training the IAF map.

In all cases, we trained the transport maps with using Adam
\citep{kingma2015adam} for 5000 steps, starting with a learning rate of 0.01,
and decaying it by a factor of 10 at step 1000 and again at step 4000. We used
a batch size of 4096 for all problems, running on a Tesla P100 GPU.

\subsubsection{HMC sampler hyperparameters}

For all HMC experiments we used the corresponding $q(\theta)$ as the initial
distribution. In all cases, we ran the 16384 chains for 1000 steps to compute
the bias and chain diagnostics, and ran with 4096 chains to compute
steps/second.

Without prior information about the target distribution, the standard practice
for tuning the HMC step size $\epsilon$ and number of leapfrog steps $L$ is to
run multiple pilot runs until acceptable behavior is observed in the chain
traces, gross chain statistics and other heuristics. For this work, we automate
this process by minimizing:
\begin{equation}
    \label{equation:optimization-objective}
    \hat{R} - \exp\left(\frac{-(\hat{R} - 1)^2}{0.02}\right) \textrm{ESS/grad},
\end{equation}
using Bayesian optimization, where ESS/grad is the effective sample size as
defined by \citet{hoffman2011no} normalized by the number of target
distribution gradient evaluations and $\hat{R}$ is potential scale reduction
\citep{gelman1992}. When computing ESS and $\hat{R}$ we use the per-component
second moment rather than the more typical mean as we are interested in how
well the HMC chains explore the tails of the target distributions. We compute
$\hat{R}$ by starting the chains from our initial distribution which, while
convenient, is not the recommended practice as it is underdispersed with
respect to the target distribution. The $\hat{R}$ values we obtain should
therefore be interpreted as lower bounds. As our distributions have multiple
components, we take the maximum $\hat{R}$ and minimum ESS/grad when computing
the optimization objective.

The intuition behind \Cref{equation:optimization-objective} is that we want the
chains to fully explore the target distribution, leading to a low $\hat{R}$.
Once that is low enough we also want the chains to be efficient by minimizing
$\textrm{ESS/grad}$. In practice, for all the samplers tested all the chain
reach an $\hat{R} < 1.1$.

When optimizing $\epsilon$ and $L$ we select the step sizes from $\epsilon \in
[10^{-4}, 5]$ and number of leapfrog steps from $L \in [1, 100]$.

\subsubsection{Results}

\begin{figure*}[!tb]
  \centering
  \includegraphics[width=0.33\textwidth]{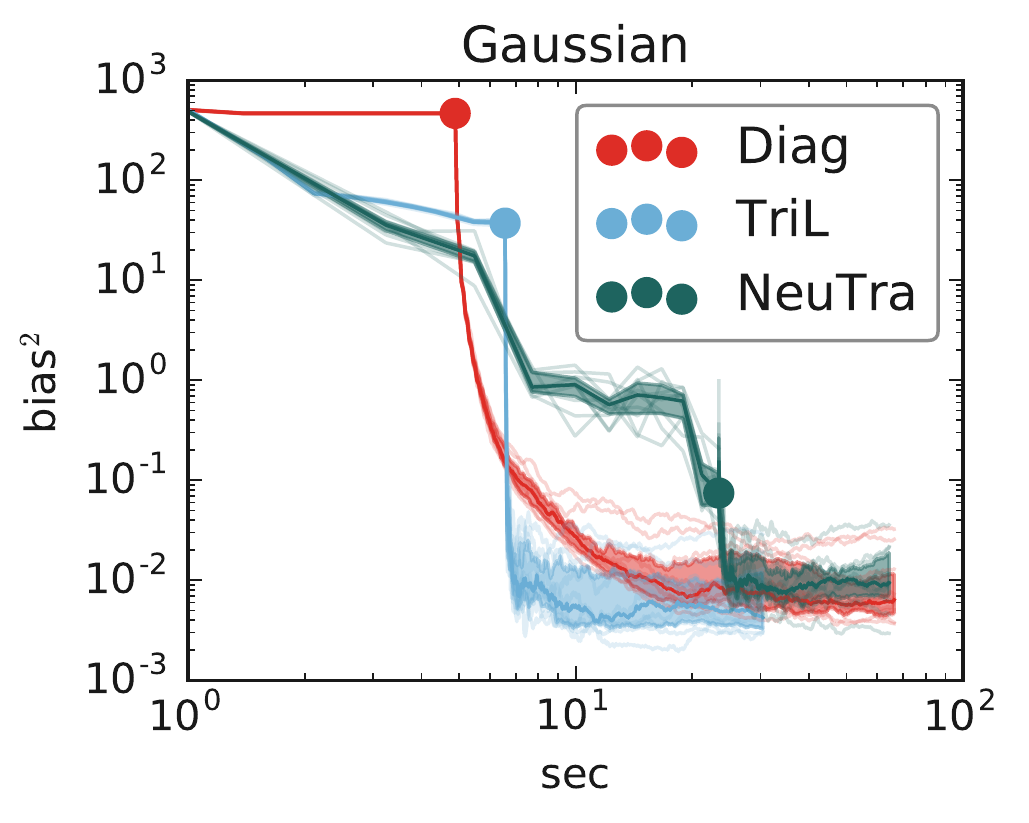}
  \includegraphics[width=0.33\textwidth]{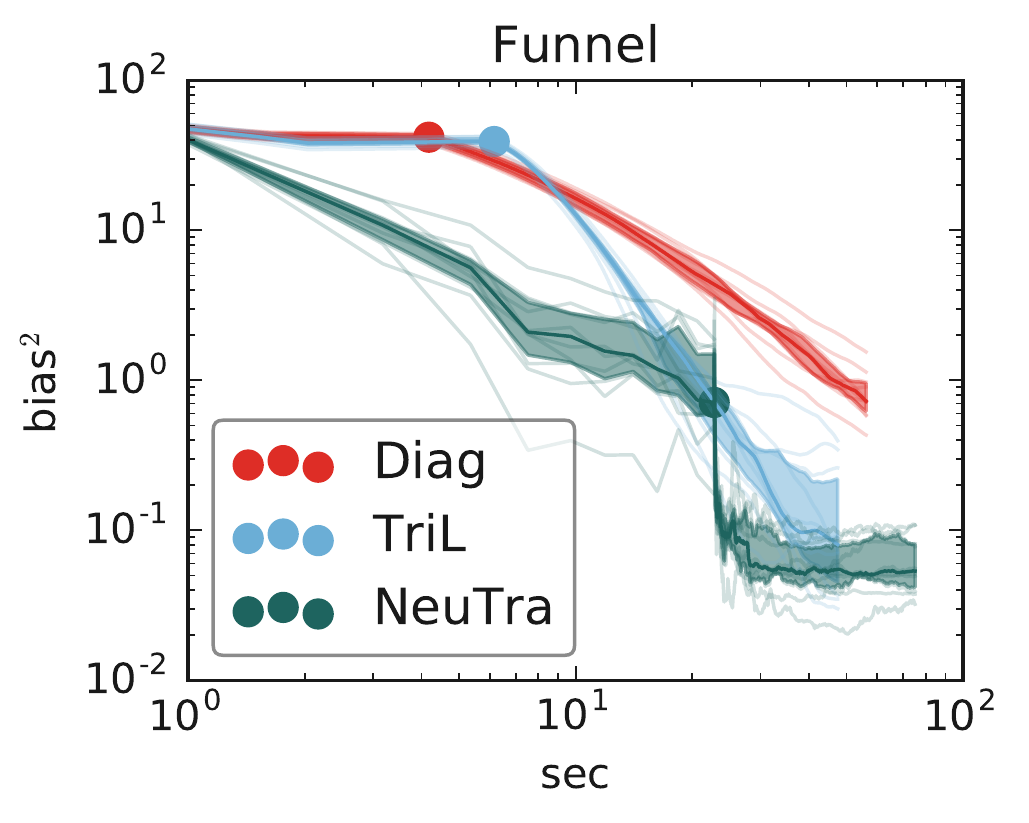}
  \includegraphics[width=0.33\textwidth]{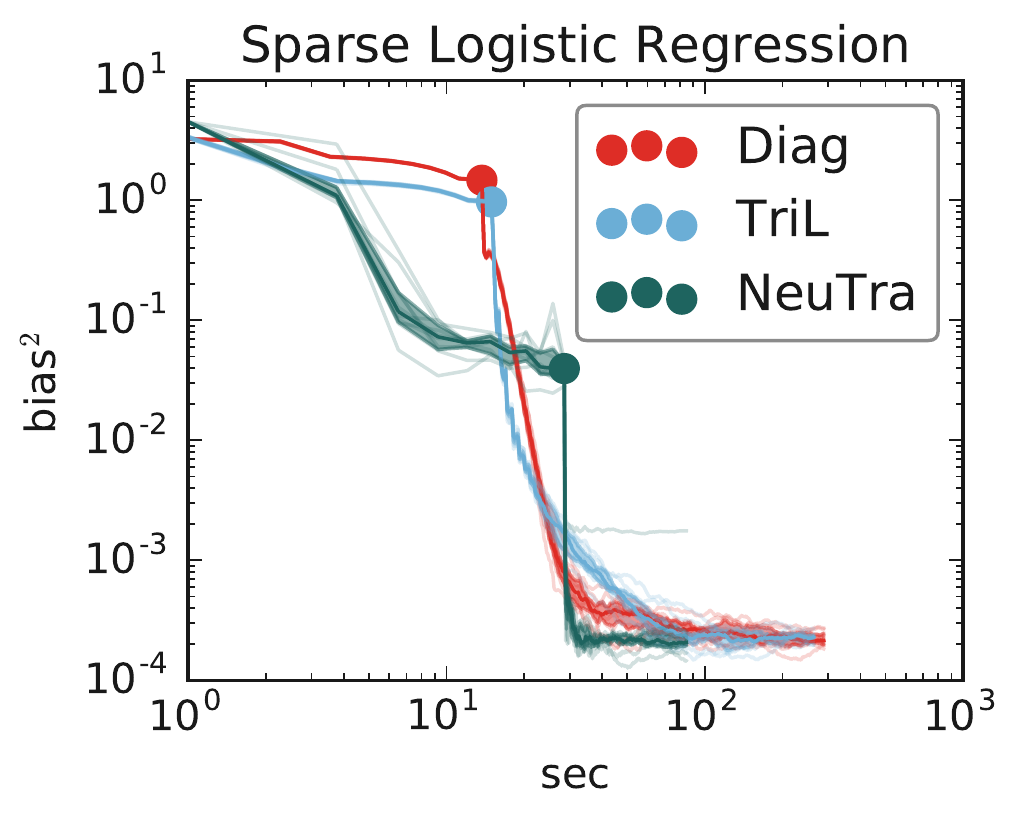}
\caption{
    Mean $\mathrm{bias}^2$ across the target distribution components vs
    wall-clock time of the variational distribution during training and HMC
    sampling. Each curve is composed of two stages: first we train the
    transport map and measure the bias of the samples from the the pushforward
    variational distribution. The end of training is marked by a circle. After
    training, we run the HMC chain for up to 1000 steps, discarding the first
    half of the chain and estimating the bias from the rest. 
    The bias estimates do not decrease indefinitely, but this is due to Monte Carlo
    noise in the estimator rather than true asymptotic bias.
    We plot the median of 10 runs (solid line), shade between the lower
    and upper quartiles and additionally show the individual runs (faint
    lines). Lower is better.
}
\label{fig:bias}
\end{figure*}

\Cref{fig:samples} shows samples from the IAF variational distribution as well
as the diagonal covariance matrix and the corresponding HMC samples. IAF
matches the target distribution very well for both Gaussian and Funnel target
distributions, with the remainder taken care of by the NeuTra HMC. The Diag
transport map does not match the target distributions all that well, but HMC
still manages to recover the target distribution due to its unbiased nature.
Note that despite this, HMC has trouble with the neck of the funnel because of
the difficult geometry in that region, while NeuTra HMC does better because the
transport map has partially simplified that region.

\begin{figure*}[!tb]
  \centering
  \includegraphics[width=0.33\textwidth]{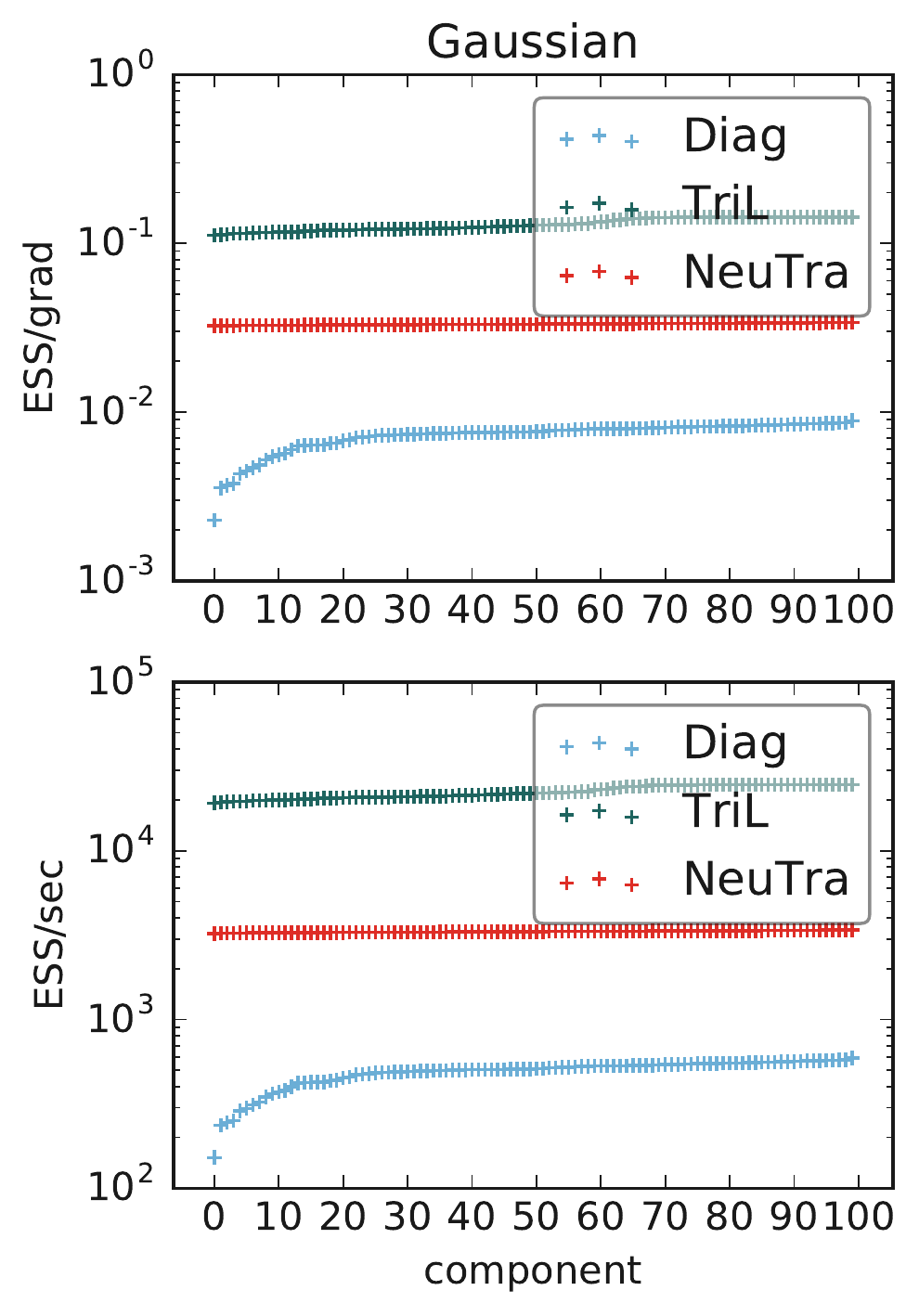}
  \includegraphics[width=0.33\textwidth]{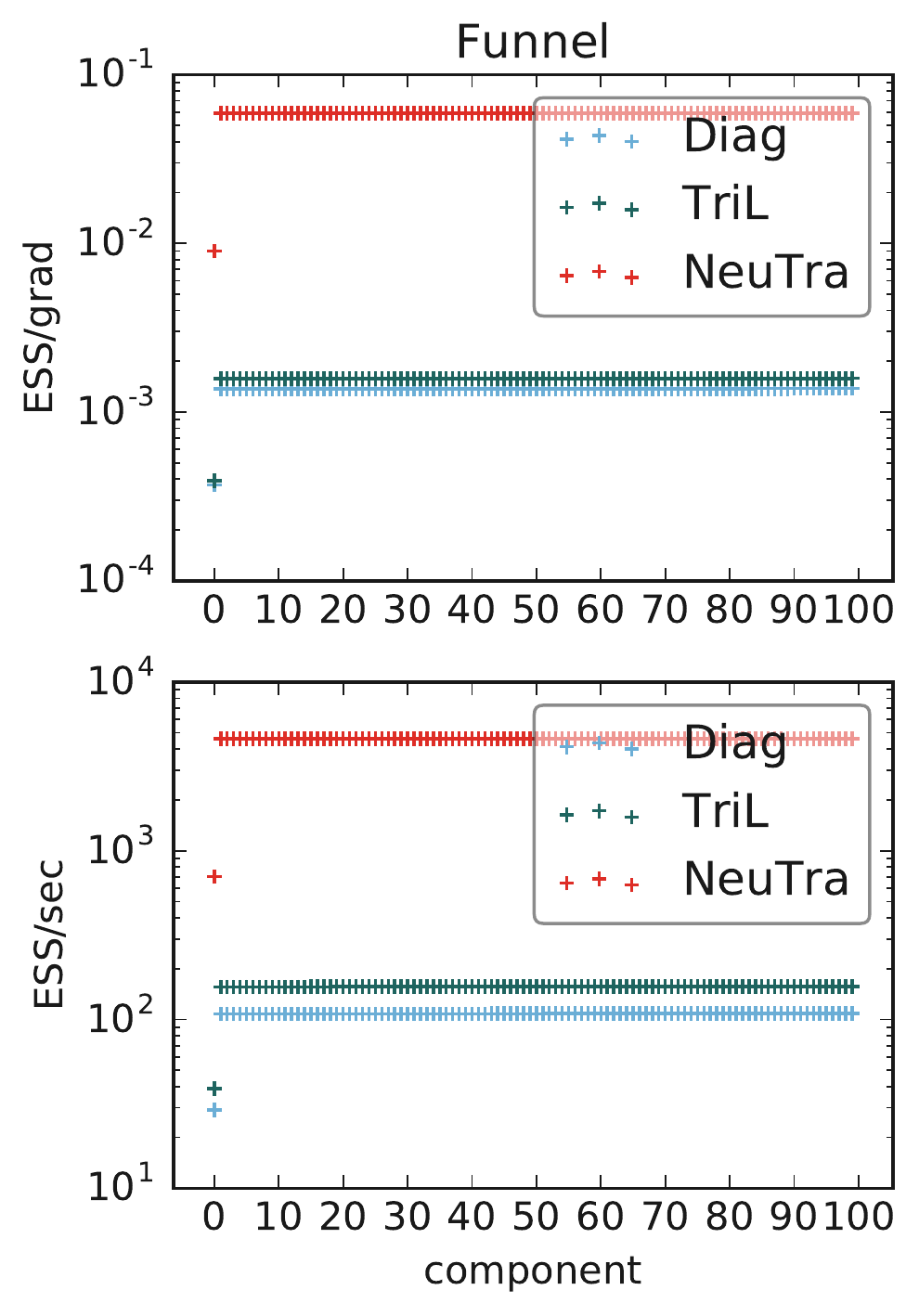}
  \includegraphics[width=0.33\textwidth]{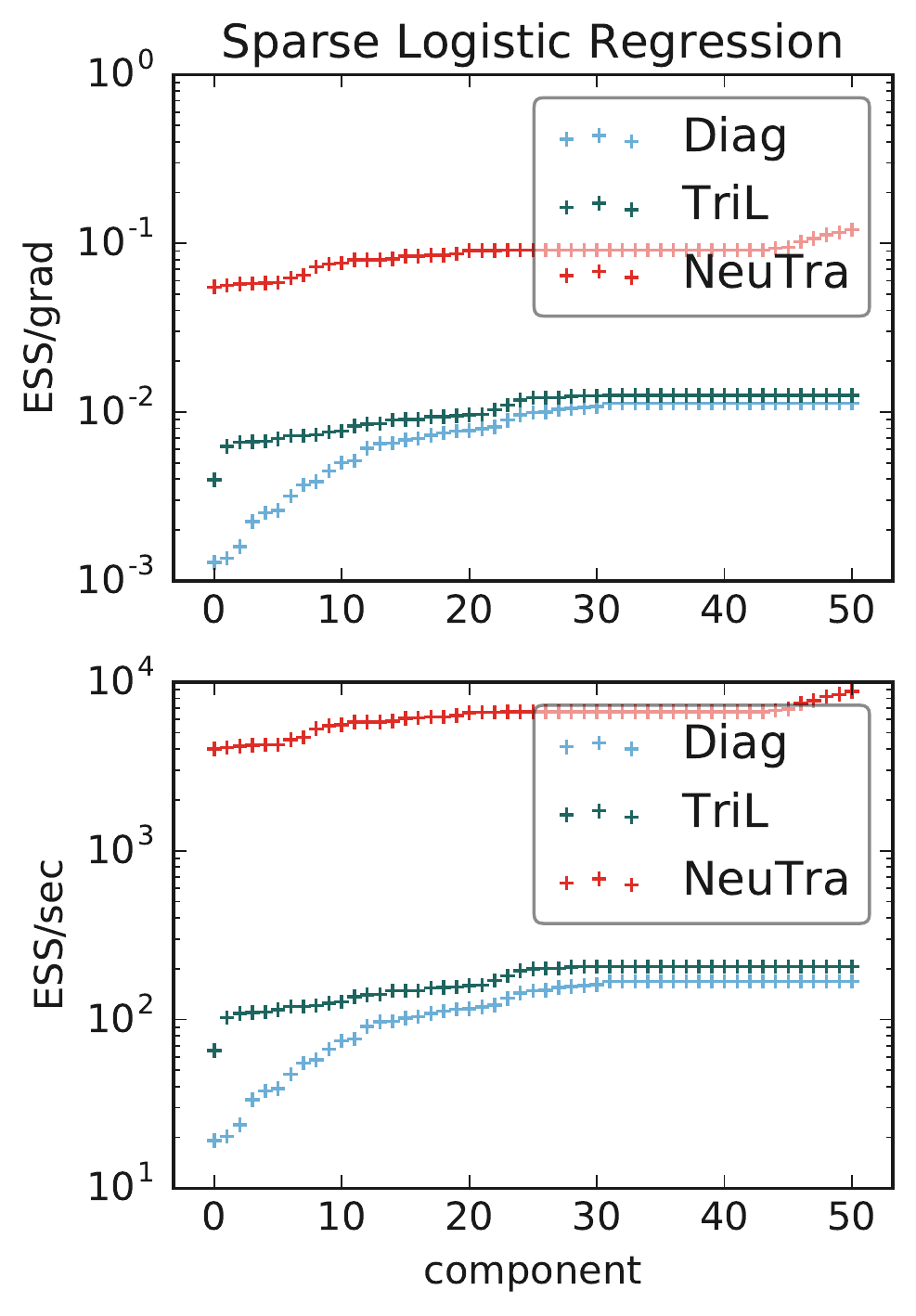}
\caption{
   Componentwise HMC chain statistics obtained after 1000 samples with first
    500 samples discarded, averaged over 10 runs (inter-run variation is too
    small to see on this plot). We sort the components based on
    the corresponding statistic value. Higher is better.
}
\label{fig:stats}
\end{figure*}

A natural concern about NeuTra HMC is that the time spent training the
neural transport map could be instead spent collecting samples from a
vanilla HMC sampler. On the other hand, NeuTra HMC might not need as
long a warmup (or ``burn-in'') period to forget its initialization,
firstly because the chains can be initialized with samples from the
variational distribution (which has been tuned to be close to the true
posterior), and secondly because the chains may mix faster after
initialization. Depending on which of these considerations
(transport-map training time versus warmup speed) dominates, either
NeuTra HMC or vanilla HMC might start generating unbiased samples
first.

\Cref{fig:bias} investigates which of these effects is dominant.
We estimated the transient squared bias
(averaged across dimensions) when estimating the second moment
of each dimension as a function of the wall-clock time.
We estimate the bias by averaging across 16384 chains (8x as
many variational samples) which gives us a noise floor due to the variance
between the chains/samples. For the chains, we discard the first half of the
chain as per the standard practice \citep{angelino2016}.

First, we train the transport map using the variational objective. This
objective is not guaranteed to reduce bias depending on the exact relationship
between the map parameterization and the target distribution, but we
nonetheless observe that it typically does. Critically, when the map is not
flexible enough to match the target distribution, estimates based on samples
from the variational distribution are biased.

After the distribution is trained, we start the HMC sampling, which
asymptotically can reduce the bias to 0 exponentially fast
\citep{angelino2016}.
For the Gaussian distribution we observe that HMC with the diagonal
preconditioner has trouble converging quickly, although by being so
computationally cheap it still overtakes NeuTra by the time it finishes
training its IAF. For this distribution the optimal preconditioner is a
TriL matrix, so it is not surprising that it performs the best.

For the Funnel the non-neural maps don't make much progress, and their
corresponding chains mix and warm up slowly. NeuTra, on the other hand reaches
the noise floor of our bias estimator quickly.

For sparse logistic regresion we observe similar behavior, although the target
distribution is well behaved enough for the non-neural transport maps to also
reach the noise floor of our bias estimator.

Another way to interpret \Cref{fig:bias} is as a practitioner's rule to decide
which algorithm to use based on their time and bias requirements. If the bias
requirements are not very stringent, the practioner may opt to use a simpler
preconditioner which can reach that target level of bias sooner. In fact, for
some problems it may be worth to forgo HMC altogether and use the samples from
the variational approximation instead, which supports the common choice of that
method in many Bayesian learning applications.

We also investigate the asymptotic behavior of the samplers by measuring the ESS
normalized by the number of gradient evaluations and wall-clock duration of a
step \Cref{fig:stats}. As before, we look at estimating the second moment of
each of the target distribution's components. In all cases except the Gaussian
NeuTra significantly outperforms the non-neural transport maps, often by over
an order of magnitude. This is a combination of two effects. First, NeuTra
simplifies the target distribution geometry, allowing HMC to explore it more
effectively. Second, HMC using non-neural transport maps needs to take many
leapfrog steps to reduce the autocorrelation, which may take more wall-clock
time than NeuTra even if NeuTra's neural transport map takes longer
per leapfrog step.

\subsection{Conditional Target Distributions}
\label{sub:conditional}

Using NeuTra HMC to generate samples from the posterior of a deep latent gaussian
model (DLGM) during training is a natural application of our technique.
Classically, DLGMs have been trained by constructing an amortized approximate
posterior, and then using variational inference to train both the approximate
posterior and generative model parameters
\citep{kingma2014autoencoding,rezende2014stochastic}. More recently, by
utilizing neural-net transport maps the quality of the approximate posterior has
been improved, yielding higher-quality generative models \citep{rezende2015normalizing}.
To incorporate NeuTra HMC into these models we build upon the interleaved
training procedure of
\citet{hoffman2017learning}. The
parameters of the approximate posterior and the transport map are trained
using the standard ELBO. For each minibatch we initialize the NeuTra HMC
chain at the sample from the approximate posterior, and then take a small
number of NeuTra HMC steps, taking the final state as the sample used to train the
generative model. We use a step size of $0.1$ and $4$ leapfrog steps.

\begin{table}
    \caption{Using NeuTra HMC to improve amortized variational inference for
    dynamically binarized MNIST. We report the test NLL averaged over 5
    separate neural net random initializations. For NeuTra HMC, the step size was
    0.1, and number of leapfrog steps was 4.}
\vskip0.1in
    \label{tb:vae}
    \centering
    \begin{tabular}{lcc}
        \toprule
            Posterior & $\log p(x)$\\
        \midrule
            Independent Gaussian & $80.84\pm0.02$\\
            IAF                  & $79.76\pm0.03$\\
            IAF+NeuTra HMC (1 step)   & $79.54\pm0.02$\\
            IAF+NeuTra HMC (2 steps)  & $79.42\pm0.02$\\
            IAF+NeuTra HMC (4 steps)  & $79.35\pm0.01$\\
        \bottomrule
    \end{tabular}
\end{table}

We use the convolutional architecture from \citet{kingma2016iaf} with the IAF
map and train it on dynamically binarized MNIST, reporting the test NLL
computed via AIS (20 chains, 10000 interpolation steps)
\citep{wu2016quantitative}. \Cref{tb:vae} shows that even a very flexible
approximate posterior can be refined via NeuTra HMC. Crucially, no new
parameters were added to the model; we simply utilized the IAF transport map
used in the standard training procedure. One caveat is that, as
reported by \citet{hoffman2017learning},
training speed is significantly reduced due to the additional evaluations of
the model for each leapfrog step.

\section{Discussion}
\label{sec:discussion}

We described Neural-Transport (NeuTra) HMC, a method for accelerating
Hamiltonian Monte Carlo sampling by nonlinearly warping the geometry
of the target distribution using inverse autoregressive flows trained
using variational inference. Using IAFs instead of affine
flows often dramatically improves mixing speed, especially on
posteriors often found in hierarchical Bayesian models.

One remaining concern is that, if the maps fail to adequately capture
the geometry of the target distribution, NeuTra could actually slow
mixing in the tails. It would be interesting to explore architectures
and regularization strategies that could safeguard against this.

\bibliography{main}
\bibliographystyle{icml2019}

\end{document}